\documentstyle[12pt]{article}
\begin{document}
\begin{center}
{\Large\bf Collective excitations in asymmetric parabolic quantum wells}

\vspace*{1cm}
{\bf P. I. Tamborenea and S. Das Sarma \\
Department of Physics \\
University of Maryland \\
College Park, Maryland  20742 }

\end{center}


\begin{center}
{\it Abstract}
\end{center}

{\it
\noindent We study within the linear response theory the far-infrared
optical absorption spectrum and
the collective excitations of an asymmetric, wide parabolic
Al$_x$Ga$_{1-x}$As quantum well, consisting of two half-parabolas of
different curvatures, in the presence of an in-plane magnetic field.
We employ a self-consistent-field approach in the local-density
approximation using parameter values of a recent
experimental study.
At low $N_s$, there is only one resonance corresponding to the lowest
inter-Landau-level transition, and at high
$N_s$ we obtain two resonances at frequencies corresponding to the
different curvatures of the well, in good agreement with experiment.
We also calculate the optical spectra without any magnetic field and
find similar trends as a function of $N_s$, i.e.,
one resonance at low $N_s$ and two main resonances at high $N_s$.}

\vfill\eject

\vspace{1cm}
Remotely doped parabolic quantum wells (PQW) show infrared magneto-optical
spectra \cite{das,kar,wix} in agreement with the generalized
Kohn theorem \cite{bre-joh-hal}, which states that an electron gas in
a perfect parabolic confinement and with a magnetic field applied in an
arbitrary direction, absorbs long-wavelength light only at the two
frequencies that correspond to excitations in the center of mass (CM)
motion---the so-called ``Kohn modes''.
For a number of confining potentials with deviations from perfect
parabolicity, both experiment \cite{wix,mag-rot} and theory
\cite{bre-dem-joh-hal,dem-tilted,sto-das1} show that the CM modes are robust
against
the nonparabolicities (although their energies may be slightly shifted), and
that additional excitations, usually with small oscillator strength,
appear in the far-infrared (FIR) optical spectra.
In the presence of an in-plane magnetic field the changes in the spectra
induced by nonparabolicities may be more dramatic, as shown for an overfilled
PQW \cite{dem-inplane}, where the CM mode, which in the perfect parabolic case
is separated from the continua of inter-Landau-level transitions,
is broadened into a continuum.

Recently \cite{yin}, the FIR optical absorption spectra of an
asymmetric PQW consisting of two half parabolas of different curvatures
have been measured in the presence of an in-plane magnetic field $\vec{B}$,
with normally incident radiation linearly polarized with
$\vec{E}^{RAD} \perp \vec{B}$.
For a perfect PQW (given by $V(z)=\alpha_0 z^2$) in this Voigt geometry,
the generalized Kohn theorem predicts a unique absorption frequency
$\omega = (\omega_0^2 + \omega_c^2)^{1/2}$,
where
$\omega_0=(2 \alpha_0 / m^\ast)^{1/2}$
is the frequency of the harmonic-oscillator confining potential
with $m^\ast$ being the electron effective mass in GaAs, and
$\omega_c=(eB/m^\ast c)^{1/2}$ is the cyclotron frequency.
In the perfect parabolic case this resonance is expected to be
independent of the electron sheet density $N_s$, as corroborated by
experiment \cite{kar,wix}.
For the asymmetric PQW, instead, the generalized Kohn theorem is no
longer valid, and the FIR spectra are seen to depend on $N_s$.
For large $N_s$ there are two resonances corresponding to the two
curvatures in the well, and as $N_s$ is decreased they merge into
a unique resonance corresponding to the lowest inter-Landau-level
transition.

In this paper, we present results of a self-consistent quantum mechanical
calculation of optical spectra and collective excitations for the asymmetric
PQW experimentally studied in Ref.~\cite{yin}, both with a magnetic field
in the Voigt geometry and without a magnetic field.
With an applied magnetic field, in the limits of low and high
$N_s$ we obtain excellent agreement with the experimental
results, and we calculate the induced density fluctuations associated with the
resonances observed at high $N_s$.
With $\vec{B}=0$ we obtain a similar behavior in the FIR spectra
as a function of $N_s$, with one resonance at low $N_s$ and two
resonances at the Kohn frequencies at high $N_s$.

Intersubband optical absorption for a quasi-2D electron gas in the
presence of an in-plane magnetic field, with the radiation polarized
in the growth direction has been calculated by Ando \cite{ando75,ando78}, and
his treatment was adapted to the normal incidence case by
Dempsey and Halperin \cite{dem-inplane}.
In this paper we use a similar self-consistent-field approach, which
includes exchange and correlations effects within the Local Density
Approximation (LDA).

We choose the growth axis in the z-direction, the magnetic field
$\vec{B}=(0,B,0)$ and the gauge $\vec{A}=(Bz,0,0)$.
Ignoring the Zeeman energy, the effective single-particle Hamiltonian
is \cite{dem-inplane,sto-das2}

\begin{equation}
H= \frac{(p_x+m^\ast\omega_cz)^2}{2m^\ast}+ \frac{p_y^2}{2m^\ast}
+ \frac{p_z^2}{2m^\ast}+V_{\mbox{eff}}(z),
\end{equation}

\bigskip
\noindent where
$V_{\mbox{eff}}(z)=V_{\mbox{conf}}+V_H(z)+V_{\mbox{xc}}(z)$
is the sum of the confining $V_{\mbox{conf}}(z)$,
Hartree  $V_H$, and exchange-correlation $V_{xc}$ \cite{ste-das} potentials.
The eigenfunctions of $H$ can be factorized as

\begin{equation}
\psi_{nk_xk_y}(x,y,z)=\frac{e^{ik_xx}}{L_x^{1/2}}\frac{e^{ik_yy}}{L_y^{1/2}}
\; \varphi_{nk_x}(z),
\end{equation}

\bigskip
\noindent and the corresponding eigenenergies are
$E_n(k_x,k_y)=\frac{\hbar^2 k_y^2}{2m^\ast}+\varepsilon_n(k_x)$.
In this work, the quasi-continuous set of values of the quantum
number $k_x$ obtained when periodic boundary conditions are applied
in the x-direction is discretized into a coarser mesh
to make the problem tractable numerically \cite{ando78,dem-inplane}.
The electron density is then calculated as

\begin{equation}
n(z)=\sum_{nk_x} N_{nk_x}  \mid \varphi_{nk_x}(z) \mid ^2,
\end{equation}

\bigskip
\noindent where $N_{nk_x}$ is given, at zero temperature, by

\begin{equation}
N_{nk_x}=\frac{2}{\pi L_x} (\frac{2m^\ast}{\hbar^2})^{1/2}\sum_{nk_x}
\theta(E_F-\varepsilon_n(k_x)) \; (E_F-\varepsilon_n(k_x))^{1/2},
\end{equation}

\bigskip
\noindent where $E_F$ is the Fermi energy and $\theta$ is the unit step
function.
The self-consistent single-particle problem without magnetic
field is solved following a method analogous to the one just
described \cite{bre-dem-joh-hal,sto-das1}.
Having solved the self-consistent single-particle problem
we proceed to calculate the absorption spectrum.
In the Voigt geometry, the absorbed power is proportional to the real
part of the xx-component of the modified two dimensional
conductivity tensor, which is related to the zz-component
by \cite{dem-inplane}
$Re[\tilde{\sigma}^{2D}_{xx}]=\frac{\omega_c^2}{\omega^2}
 Re[\tilde{\sigma}^{2D}_{zz}]$.
In our calculations with magnetic field we obtain $\tilde{\sigma}^{2D}_{zz}$
following the method described in Refs. \cite{ando78,dem-inplane},
and for the zero magnetic field calculations we follow Refs.
\cite{ando77,bre-dem-joh-hal,sto-das1}.
In both cases one obtains

\begin{equation}
\tilde{\sigma}^{2D}_{zz}=-i\omega e^2 \sum_\eta \frac{f_\eta^{(z)}}
{\tilde{\varepsilon}_\eta^2-(\hbar\omega)^2-2i\hbar^2\omega/\tau},
\end{equation}

\noindent where $f_\eta^{(z)}$ are the oscillator strengths corresponding to
the
resonant energies $\tilde{\varepsilon}_\eta$, which include the
depolarization and the exciton-like corrections \cite{ando78,ando77},
and $\tau$ is a phenomenological relaxation time.
In the Voigt geometry problem the index $\eta$ denotes a discretized
quasi-continuous variable whereas in the zero magnetic field case it
is simply an integer variable.
This implies that at zero magnetic field the absorption spectrum
consists of a set of isolated resonant frequencies, and with an in-plane
magnetic field there can be, in principle, both isolated and broad
quasi-continuum resonances, which is indeed the case for the
asymmetric parabolic well under consideration.

Our model potential describing the asymmetric PQW sample studied in
Ref. \cite{yin} is shown in Fig.~1,
and consists of two half-parabolas with curvatures
$\alpha_1=5.1 \times 10^{-5} \; meV/\mbox{\AA}^2$ and
$\alpha_2=6.2 \times 10^{-5} \; meV/\mbox{\AA}^2$,
and it is $3000 \; \mbox{\AA}$ wide.
For the calculations in the Voigt geometry, we choose a magnetic
field in the region of the observed resonances of $B=5.8 \; T$.
In Fig.~1 we show the calculated self-consistent potentials
for some values of the sheet density $N_s$, which, as expected, become
flattened over a wider region in the center of the well, as the electron
slab width increases with $N_s$.
Note that for the values of $N_s$ considered in the experiment, and in
our calculations with magnetic field, only the lowest subband
is occupied at zero temperature.
The electron density profiles $n(z)$ are shown as an inset in Fig.~1.
We checked that the abrupt change in the curvature at $z=0$ does not produce
unphysical results by using a similar potential with a graded change
in curvature, and the density profile was identical to the original
one within our numerical precision.

The optical absorption spectra for the asymmetric quantum well
and for several densities $N_s$ in the range used in Ref. \cite{yin} are
shown in Fig.~2.
The frequencies associated with the half-parabolas of curvatures
$\alpha_{1,2}$ are $\omega_{1,2}=2\alpha_{1,2}/m^\ast$.
The frequencies $w_{K1,K2}$ marked for reference on each panel of Fig.~2
are the ``Kohn'' frequencies of each half-parabola,
$w_{K1,K2}= (\omega_{1,2}^2+\omega_c^2)^{1/2}$---the only frequencies at which
perfect parabolic wells with curvatures $\alpha_{1,2}$ would absorb
long-wavelength radiation in the Voigt geometry.
The phenomenological scattering time $\tau=20 \times 10^{-12} s$,
taken to be a constant in our calculations, is taken
from the linewidth of the narrower resonance in the experiment \cite{yin}.
For a perfect PQW the generalized Kohn theorem predicts a unique
resonance independent of the electron density.
In our case, in contrast, we obtain a strong dependence on $N_s$.
For small $N_s \stackrel{<}{\sim} 4.7\times10^{10} \; cm ^{-2}$
there is only one resonance (for example, $\tilde{\varepsilon}=10.644 \; meV$
at
$N_s=0.1 \times 10^{10} \; cm ^{-2}$),
corresponding to the optical transition between the first two energy
levels of the bare confining potential in the presence of the in-plane
magnetic field, with an energy
$\varepsilon_1(k_x=0)-\varepsilon_0(k_x=0)=10.648 \; meV$.
As $N_s$ increases another resonance appears at an energy lower than
$\omega_{K1}$, and gradually moves to $\omega_{K1}$ while the other
resonance also shifts to end up at $\omega_{K2}$ for high $N_s$.
At a sheet density $N_s=7.5 \times 10^{10} \; cm ^{-2}$ (the highest
$N_s$ reported in the experiment) we obtain two resonances
$\omega_{1,2}^{RES}$ very close to the ``Kohn'' frequencies
$\omega_{K1,K2}$.
We note that a direct comparison of the theoretical and calculated
spectra is not possible because the experimental results correspond
to a magnetic field sweep at a constant resonance frequency whereas
the calculated optical spectra are given as a function of frequency
at a fixed magnetic field.
In order to make a quantitative comparison with experiment, we
use the resonant frequencies $\omega_{1,2}^{RES}$ to compute the
harmonic oscillator frequencies of the well halves
$\omega_{1,2}^{CALC}=((\omega_{1,2}^{RES})^2-\omega_c^2)^{1/2}$,
and obtain their difference
$\omega_{2}^{CALC}-\omega_{1}^{CALC} = 0.378 \; meV$.
This shows an 8\% discrepancy with the input value
$\omega_{2}-\omega_{1} = 0.350 \; meV$ ($\omega_{1,2}$
are the values used in the definition of the confining potential)
which is comparable to the 4\% discrepancy found in experiment
at the same sheet density.
Therefore, we obtain a very good agreement with the experimental
results of Ref.~\cite{yin} for low and high $N_s$.
The agreement at high $N_s$ also confirms the strong magnetic field
classical arguments based on the Magarill-Chaplik \cite{mag-cha}
theory given in Ref.~\cite{yin}.
At intermediate $N_s$, however, the position of the weaker resonance
seems to evolve in different ways as a function of $N_s$ in theory
and experiment.
In the experiment, the two ``Kohn'' resonances found at high $N_s$
appear to merge into one absorption peak as $N_s$ is reduced, in
contrast to the separation and vanishing of one of the resonances found
in our calculation.
One possibility is that this discrepancy appears as a result of the
different type of spectra obtained in theory and experiment (frequency
versus magnetic field sweep) and then an improved fit to the experimental
results at intermediate $N_s$ could be obtained if a magnetic field and
density dependent scattering time $\tau(B,N_s)$ is included in the
calculation.

Now we comment on two aspects of the collective excitations
of the electron gas in the asymmetric parabolic well in the
presence of an in-plane magnetic field.
The CM mode predicted by the generalized Kohn theorem for a
perfect parabolic confinement is a rigid oscillatory motion
of the electron slab. In linear response, its density fluctuations
$\delta n(z)$ are proportional to $dn(z)/dz$ \cite{dem-inplane}.
In Fig.~3 we show the fluctuation profiles $\delta n(z)$ for
the resonant modes at $N_s=7.5 \times 10^{10} \; cm^{-2}$
together with the derivative $dn(z)/dz$.
We see that the resonant modes very accurately describe
rigid translations of the electron gas in the corresponding
halves of the well---right and left halves for the resonances
with $\omega_{K1}$ and $\omega_{K2}$, respectively.
Finally, a notable difference between our results for
the asymmetric PQW of Ref. \cite{yin} and a previous study of an
overfilled PQW in the Voigt geometry \cite{dem-inplane} is the interplay
between the main resonances or ``Kohn modes'' and the continuum of
inter-Landau-level transitions.
For the overfilled PQW the CM mode
is broadened into a continuum, whereas for the asymmetric PQW
the ``Kohn modes'' of the two half parabolas are split off from the
continuum of inter-Landau-levels at high densities, and at lower densities
one of them becomes a broad continuum resonance while the other one
remains an isolated resonance as in the case of perfect parabolic
confinement.

Figure 4 shows the calculated IR optical spectra for the asymmetric
PQW without magnetic field for $N_s$ in the range
$0.1-1.6 \times 10^{11} \; cm^{-2}$.
The number of occupied subbands goes from 1 to 4, and we keep
10 subbands in the calculation of the optical spectra.
Again, for low $N_s$ there is only one resonance corresponding
to the lowest optical transition of the bare well, and at high
$N_s$ there are two resonances close to the frequencies
$\omega_1$ and $\omega_2$.
A surprising feature is the appearance of two resonances
of similar weight at $N_s=0.6 \times 10^{11} \; cm^{-2}$,
when the second subband is populated, which disappears
at higher $N_s$.

In conclusion, we have studied the FIR optical absorption spectra
of an asymmetric PQW with a magnetic field in the Voigt geometry
and without a magnetic field using the self-consistent LDA approach.
We compare our magnetic field results with experimental spectra and
find good quantitative agreement with experiment for low and high
$N_s$ and quantitative differences in the spectra at
intermediate $N_s$.
The origin of this disagreement remains an open question.
In particular, experiments with a frequency sweep at constant
magnetic field would permit a more direct comparison between
theory an experiment.
At zero magnetic field, the general trends of the spectra are similar
to those of the Voigt geometry results, and here again an experimental
measurement of the optical spectra would be useful to compare
the accuracy of the LDA approximation with and without
magnetic field.

The authors are grateful to I. K. Marmorkos, Y. K. Hu, K. Karrai,
C. Stafford, and especially to H. D. Drew for useful discussions.
This work is supported by the US-ARO, US-ONR, and the US-DoD.

\noindent{\bf Figure Captions}

\noindent {Figure 1}

\noindent Model bare potential of the asymmetric parabolic well (solid line),
and calculated self-consistent potentials with an in-plane magnetic
field of $B=5.8 \; T$, for $N_s=2.4$ (dash line),
4.7 (dotted line), and $7.5  \times 10^{10} \; cm ^{-2}$ (dash-dot line);
inset: corresponding calculated self-consistent densities.

\vspace*{.2in}
\noindent {Figure 2}

\noindent Calculated absorption spectra for the asymmetric parabolic
quantum well
with a magnetic field $B=5.8 \; T$ in the Voigt geometry for various sheet
densities $N_s$ (given in units of $10^{10} \; cm^{-2}$).
$\omega_{K1}$ and $\omega_{K2}$ are the ``Kohn'' frequencies associated
with the curvatures of the two half-parabolas of the well.

\vspace*{.2in}
\noindent {Figure 3}

\noindent Density fluctuations $\delta n(z)$ (dotted lines) associated
with the
two absorption peaks at $N_s=7.5 \times 10^{10} \; cm ^{-2}$ shown
in Fig.~2, and the derivative of the ground-state electron
density $dn(z)/dz$ (solid lines).
The partial agreements between $\delta n(z)$ and $dn(z)/dz$
show that the density fluctuations of each mode correspond to rigid
oscillations of the electron slab on each half of the asymmetric well,
in analogy to the center-of-mass mode present in perfect parabolic wells.

\vspace*{.2in}
\noindent {Figure 4}

\noindent Calculated absorption spectra for the asymmetric parabolic
quantum well without magnetic field for various sheet densities $N_s$.
At high $N_s$ the resonances agree with the curvatures of the well,
$\omega_1=3.41 \; meV$ and $\omega_2=3.76 \; meV$.

\end{document}